\documentclass[11pt]{article}

\usepackage[preprint]{acl}

\usepackage{times}
\usepackage{latexsym}
\usepackage{booktabs}
\usepackage{amsmath}
\newenvironment{formal}
{\begin{quote}\em}
{\end{quote}}
\usepackage{multirow}

\usepackage[T1]{fontenc}
\usepackage[utf8]{inputenc}
\usepackage{enumitem}
\usepackage{microtype}

\usepackage{inconsolata}

\usepackage{graphicx}
\usepackage{colortbl}
\definecolor{bestbg}{RGB}
{255,235,235}   % 淡红
\definecolor{secondbg}{RGB}{235,245,255} % 淡蓝
\title{FinCall-Surprise: A Large Scale Multi-modal Benchmark for \\Earning Surprise Prediction}

\author{Dong Shu\textsuperscript{1}, Yanguang Liu\textsuperscript{3}, Huopu Zhang\textsuperscript{2}, \textbf{Mengnan Du\textsuperscript{3,*}}\\
\textsuperscript{1}Northwestern University \,
\textsuperscript{2}Georgia Institute of Technology \,
\textsuperscript{3}NJIT\\
\small\texttt{dongshu2024@u.northwestern.edu}, \small\texttt{hzhang931@gatech.edu}, 
\small\texttt{\{yanguang.liu,mengnan.du\}@njit.edu}\\[2pt]
\small\textsuperscript{*}Corresponding author
}

\begin{document}
\maketitle
\begin{abstract}

Predicting corporate earnings surprises is a profitable yet challenging task, as accurate forecasts can inform significant investment decisions. However, progress in this domain has been constrained by a reliance on expensive, proprietary, and text-only data, limiting the development of advanced models. To address this gap, we introduce \textbf{FinCall-Surprise} (Financial Conference Call for Earning Surprise Prediction), the first large-scale, open-source, and multi-modal dataset for earnings surprise prediction. Comprising 2,688 unique corporate conference calls from 2019 to 2021, our dataset features word-to-word conference call textual transcripts, full audio recordings, and corresponding presentation slides. We establish a comprehensive benchmark by evaluating 26 state-of-the-art unimodal and multi-modal LLMs. Our findings reveal that (1) while many models achieve high accuracy, this performance is often an illusion caused by significant class imbalance in the real-world data. (2) Some specialized financial models demonstrate unexpected weaknesses in instruction-following and language generation. (3) Although incorporating audio and visual modalities provides some performance gains, current models still struggle to leverage these signals effectively. These results highlight critical limitations in the financial reasoning capabilities of existing LLMs and establish a challenging new baseline for future research. The FinCall-Surprise dataset is available at \url{https://github.com/Tizzzzy/FinCall-Surprise}.

\end{abstract}

\section{Introduction}

In the finance domain, a conference call, often referred to as an earnings call, serves as a critical communication channel between the management of a public company and its stakeholders, including analysts, investors, and the media \cite{kimbrough2005effect}. During these calls, executives present the firm’s financial results, discuss recent performance, and provide forward-looking guidance. Analysts and investors closely analyze this information to predict their expectations of the company’s earnings per share (EPS) \cite{patell1976corporate}. An \emph{earnings surprise} occurs when the company's reported EPS differs from market expectations \cite{latane1979standardized}. Historically, a positive earnings surprise, where actual earnings exceed the reported EPS, often correlates with a subsequent increase in the company's stock price, while a negative surprise typically leads to a decline. Therefore, the ability to accurately predict an earnings surprise ahead of the official announcement is a significant challenge and an opportunity for investors seeking to inform their trading decisions \cite{skinner2002earnings}.

Given the potential financial rewards, researchers and investors have long sought to systematically analyze conference calls to predict earnings surprises. This pursuit began with traditional machine learning models \cite{dhar2001comparison, doyle2006extreme, bissessur2016analyst}, and has recently shifted attention toward large language models (LLMs) \cite{zhang2025sae, zhang2025cross, zhu2025post, liang2025does}. However, while the performance of these models is often impressive, their capabilities are inherently constrained by a fundamental limitation, which is the unimodal and text-only nature of the datasets they are trained on. In reality, human analysts do not just analyze what executives say, but also how they say it. The vocal tone and rhythm of a speaker can reveal confidence or uncertainty that is lost in a plain transcript. Similarly, visual information, such as presentation slides, provides an essential context that is often referenced during the discussion.

To address this critical gap, we introduce \textbf{FinCall-Surprise}, a novel, large-scale, multi-modal dataset specifically designed for the task of earnings surprise prediction. The dataset is composed of 2,688 unique corporate conference calls spanning from 2019 to 2021, with 919 calls from 2019, 704 from 2020, and 1,065 from 2021. Most importantly, each call in the dataset is complete, containing three synchronized modalities: 1) the word-to-word textual transcript, 2) the full audio recording of the call, and 3) the corresponding presentation slides referenced by executives. With the introduction of this dataset, our primary contributions are as follows:
\vspace{-5pt}
\begin{itemize}[leftmargin=*]\setlength\itemsep{-0.3em}
    \item We present FinCall-Surprise, the first large-scale, multi-modality dataset focused on real-world corporate earnings calls, providing a rich resource for developing and evaluating earning surprise prediction models.
    \item We establish a comprehensive benchmark by evaluating more than 20 state-of-the-art unimodal and multi-modal LLMs. 
    
    \item Our benchmarking findings reveal that current models struggle to effectively leverage the multi-modal signals for this task, establishing a challenging baseline and highlighting the need for more sophisticated architectures.
\end{itemize}

\begin{figure*}[ht]
    \centering
    \includegraphics[width=0.95\linewidth]{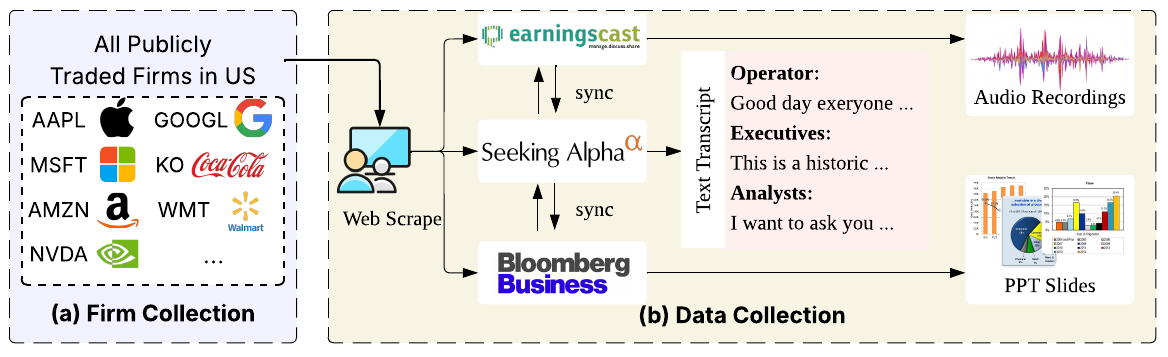}
    \caption{Overview of our data construction pipeline, which consists of two stages: \textbf{(a) Firm Collection (left):} We select large, publicly traded US firms based on market capitalization ($ > \$1$B) and daily trading volume ($ > \$50$M). \textbf{(b) Data Collection (right):} For each firm, we gather and synchronize three modalities for each quarterly earnings call: textual transcripts from Seeking Alpha, audio recordings from EarningsCast, and presentation slides from sources like Bloomberg Business.}
    \label{fig:pipeline}
\end{figure*}

\section{Related Work}

\subsection{Earning Surprise Prediction Models}

The use of AI in predicting earnings surprises has a long history that began with traditional statistical and econometric models. In early foundational research, researchers like \citet{numbers1968empirical} relied on simple models such as linear regressions. A pivotal shift occurred when \citet{ou1989financial} used a broader set of financial statement data with a logit model to predict the direction of earnings, foreshadowing the machine learning era. From approximately 2000 to 2015, classic machine learning models like SVMs, Random Forests, and Gradient Boosting took center stage \cite{dhar2001comparison, skinner2002earnings, zolotoy2012earnings}. Their key advantage was the ability to analyze hundreds of financial variables simultaneously, capturing complex, non-linear relationships that eluded older methods. The period after 2015 saw the rise of deep learning, particularly RNNs and LSTMs, which introduced unstructured text as a new and powerful data source \cite{vargas2017deep, Meursault_Liang_Routledge_Scanlon_2023, article}.

The contemporary approach to earnings surprise prediction is dominated by LLMs, which represent a revolutionary leap in analytical capability. Researchers began feeding models text from conference call transcripts and financial reports to extract sentiment and forward-looking statements \cite{koval2023forecasting}. Since LLMs can possess a deep and nuanced understanding of financial jargon and context, they can dissect conference call transcripts, press releases, and news with unprecedented accuracy \cite{araci2019finbert, kim2024financial, lee2025large}. However, due to the lack of high-quality multimodal conference call datasets, models for earnings surprise prediction have not yet reached the multimodal level.

\subsection{Earning Surprise Prediction Datasets}

Due to the high value and sensitivity of corporate financial data, most resources are stored within major commercial platforms such as WRDS, Bloomberg, and Refinitiv, which require paid subscriptions or API access. Despite this barrier, there has been steady progress in developing open-source financial datasets, each typically targeting narrower and more specialized tasks in the finance domain to advance AI research in this field. For instance, datasets such as \cite{chen2021finqa, chen2024fintextqa, zhu2021tat, lai2024sec, reddy2024docfinqa, yuan2024finllms} focus on question answering over financial reports. Other datasets have been developed for stock market prediction \cite{dong2024fnspid, Rao_NIFTY50_Kaggle, Patel_Google_Stock_Kaggle}, fraud detection \cite{feng2023empowering}, sentiment analysis \cite{Borhani_Twitter_Sentiment_Kaggle, cortis-etal-2017-semeval}, and misinformation classification \cite{10.1145/3701716.3715292}. More recent efforts have extended financial datasets into the multimodal space, motivated by both the limitations of single-modality data and the growing capabilities of LLMs to understand information beyond text \cite{li2020maec, shu2025finchart, luo2025finmme}.

Earnings surprise prediction has recently gained increasing attention in the context of LLMs, with several studies proposing new methods to improve performance on this task \cite{zhu2025post, zhang2025sae}. However, relatively few works have focused on dataset construction. Early efforts such as \citet{koval2023forecasting, koval2024financial} introduced conference call transcript datasets for this purpose, but these resources are not fully open source, limiting their accessibility and impact. Consequently, many current studies continue to depend on expensive commercial data providers, often accessing conference call content through paid APIs \cite{heater2025winning, alsabah2025love, zhang2025cross}.

Our proposed FinCall-Surprise dataset addresses this challenge by introducing the first fully open-source, multi-modal conference call dataset specifically designed for earnings surprise prediction, covering the period from 2019 to 2021. Each data in our dataset provides synchronized modalities: the complete text transcript of the conference call, the full audio recording, and the accompanying presentation slides referenced by executives. By releasing this dataset, we aim to remove the reliance on commercial APIs and enable the research community to explore earnings surprise prediction at a multimodal level, opening new directions for both financial NLP and multimodal learning.

\section{FinCall-Surprise Construction}

In this section, we detail the creation process for the FinCall-Surprise dataset.

\subsection{Data Construction Pipeline}

The construction of  FinCall-Surprise follows a comprehensive two-stage pipeline designed to collect, synchronize, and annotate data from multiple sources. Our objective is to build a robust benchmark that integrates the textual, auditory, and visual dimensions of corporate conference calls. As illustrated in Figure~\ref{fig:pipeline}, the first stage involves assembling the universe of all publicly traded firms in the United States with substantial size and liquidity. To focus on economically significant firms, we restrict the sample to companies with market capitalizations exceeding \$1 billion and average daily trading volumes above \$50 million. This initial firm collection yielded an initial pool of more than 4,000 unique companies. 

In the second stage, we acquire the three distinct modalities for each firm’s earnings conference calls. For textual transcripts, we systematically web scrape content from established financial platforms, primarily Seeking Alpha\footnote{\url{https://seekingalpha.com/}}. These transcripts are well structured, containing both speaker identifiers (operators, executives and analysts) and the corresponding speech content. For the auditory component, we web scrape the associated audio recordings from EarningsCast\footnote{\url{https://earningscast.com/}}. Finally, we construct a dataset of corporate presentation slides by collecting files from multiple sources, including Bloomberg News and company websites. Most publicly traded firms host one conference call per fiscal quarter. To ensure temporal consistency across modalities, we align all data sources by quarterly reporting periods. In addition, we use the conference call titles to cross-check and verify that the transcript, audio recording, and presentation slides correspond to the same event, ensuring that our multi-modal dataset is accurately matched at the event level.

\begin{table*}[ht]
\centering
\scalebox{0.8}{
\begin{tabular}{l|ccc|ccc|ccc|cc|c}
\toprule
& \multicolumn{3}{c|}{Transcript (words)} & \multicolumn{3}{c|}{Slide (pages)} & \multicolumn{3}{c|}{Audio (sec)} & \multicolumn{2}{c|}{Label (Percentage)} &  \\ 
\cmidrule(lr){2-4} \cmidrule(lr){5-7} \cmidrule(lr){8-10} \cmidrule(lr){11-12}
Year & Mini & Max & Avg. & Mini & Max & Avg. & Mini & Max & Avg. & Positive & Negative & Total \\ 
\midrule
2019 & 905 & 57,485 & 8,634.66 & 1 & 156 & 26.05 & 1096.07 & 8870.97 & 3642.66 & 0.79 & 0.21 & 919 \\
2020 & 865 & 46,623 & 8,925.10 & 1 & 170 & 27.80 & 1404.06 & 7585.31 & 3859.25 & 0.86 & 0.14 & 704 \\
2021 & 752 & 45,632 & 8,646.77 & 1 & 113 & 27.46 & 1197.95 & 11938.90 & 3720.29 & 0.89 & 0.11 & 1065 \\
\bottomrule
\end{tabular}}
\caption{Data statistics of the FinCall-Surprise. The dataset spans three years (2019–2021), with each conference call containing three synchronized modalities: text transcripts, presentation slides, and audio recordings. We report the minimum, maximum, and average values for each modality. Transcript length is measured in words, slides in pages, and audio in seconds. For labels, we present the percentage distribution of positive and negative classes. The last column shows how much data we have for each year.}
\label{tab:data_analysis}
\end{table*}

\subsection{Earning Surprise Label Preliminaries}

% \textcolor{red}{TODO: too similar with the SAE-fire Alice: have paraphrased this part}

Before classifying each conference call label as positive or negative, we first collected the reported Earnings Per Share (EPS) and the analyst consensus EPS forecasts from the IBES database. Following \citet{latane1979standardized}, we measure the earnings surprise (ES) using the Standardized Unexpected Earnings (SUE), defined as the difference between the reported EPS and the analyst consensus estimate, scaled by the standard deviation of analyst forecasts. The consensus estimate is calculated as the mean of the most recent valid analyst forecasts issued within one month after the conference call, allowing analysts to revise their expectations based on the call content and recent financial disclosures. This design provides a forward-looking measure of market expectations and yields a more realistic, yet challenging, prediction task. The average time span between the input transcript and the target earnings event is about three months, further highlighting the difficulty of the task.
% 需在导言区调用 amsmath 宏包: \usepackage{amsmath}
\begin{equation}
\text{ES} = \frac{\text{EPS}_{\text{reported}} - \operatorname{Avg}(\text{EPS}_{\text{estimated}})}{\operatorname{Std}(\text{EPS}_{\text{estimated}})}
\end{equation}
\begin{equation}
y =
\begin{cases}
0, & \text{ES} \leq -\delta \\
1, & \text{ES} \geq \delta
\end{cases}
\end{equation}
We convert the continuous earnings surprise (ES) into a binary classification task by assigning a label of Positive (\( +1 \)) when \( \text{ES} > \delta \) and Negative (\( 0 \)) when \( \text{ES} < -\delta \), where \( \delta = 0.50 \). This threshold follows prior studies on standardized unexpected earnings \cite{eli_bartov_patterns_1992} and price momentum \cite{luo_retail_2022}, which classify earnings surprises as large when \(|\text{SUE}| \geq 0.5\). The chosen cutoff balances sample size with event significance. Observations with immaterial surprises (i.e., \( \text{ES} \in [-0.50, 0.50] \)) are excluded, as these near-zero values typically elicit weak market responses and may reflect earnings management. Although \( ES \) is continuous, market reactions are largely binary, responding more to the direction than to the magnitude of the surprise. We therefore focus on material surprises that are more likely to influence investor behavior and asset pricing.

% \begin{figure}
%     \centering
%     \includegraphics[width=0.9\linewidth]{transcript_example.pdf}
%     \caption{An example of the conference call text transcript and the corresponding earning surprise label.}
%     \label{fig:transcript_example}
% \end{figure}

\subsection{Data Analysis}

\begin{figure*}[t]
    \centering
    \includegraphics[width=0.90\linewidth]{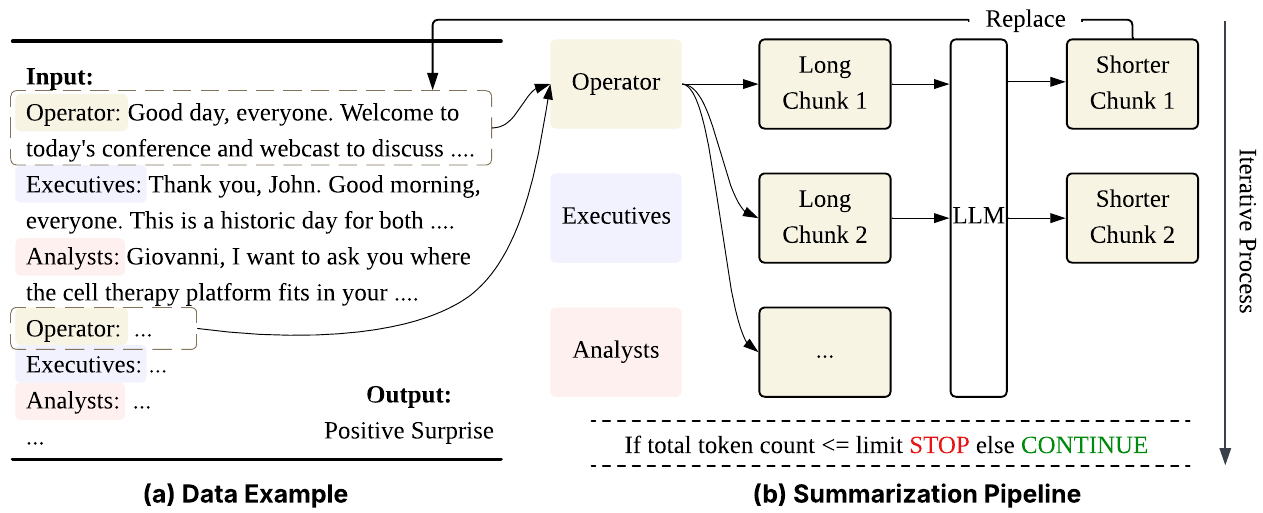}
    \caption{Illustration of our summarization pipeline. (a) A data example showing a conference call transcript with speaker turns (Operator, Executives, Analysts) and its corresponding earning surprise label. (b) The summarization pipeline, where transcripts are grouped by speaker and iteratively summarized by the LLM until the total token count falls below the predefined limit.}
    \label{fig:summarization}
\end{figure*}

A detailed statistical analysis of our dataset is presented in Table \ref{tab:data_analysis}. The statistics highlight the substantial scale and complexity of the data across all three modalities. On average, the textual transcripts for each conference call contain approximately 8,600 words. The accompanying presentation slides average 27 pages in length. Finally, the audio recordings have an average duration of approximately 3,700 seconds. Each data point in our dataset is annotated with a binary label indicating either a positive or negative earnings surprise. As shown in the last column, the distribution of labels is highly imbalanced, with the proportion of positive labels significantly exceeding that of negative labels. This imbalance is unavoidable, since our dataset is constructed from real-world financial data, where positive earnings surprises occur more frequently.
The content of the conference calls is consistently structured around three primary speaker roles: Operators, Executives, and Analysts. An  example is provided in Figure \ref{fig:summarization}. The Operator typically opens and closes the call. The Executives, including chief members such as the CEO and CFO, present the company's financial results and strategic outlook. Finally, the Analysts, representing investment firms and financial institutions, pose questions to the executives to gain deeper insights into the company's performance.

\section{Benchmark Setting}

\subsection{Baseline LLMs}

To establish a comprehensive performance benchmark for FinCall-Surprise, we selected a diverse set of 26 state-of-the-art, open- and close-source models. For a structured analysis, these models were categorized into four distinct groups.

The first group consists of general purpose unimodal models, which process text-only inputs, and includes GPT-oss-20B \cite{openai2025gptoss120bgptoss20bmodel}, Qwen-2.5-14B \cite{qwen2.5}, Mistral-7B \cite{jiang2023diego}, LLaMa-3.2-3B \cite{grattafiori2024llama}, Gemma-3-1B \cite{gemma_2025}. The second group consists of finance-specialized unimodal models fine-tuned for the financial domain, including Finance-LLaMa3-8B \cite{cheng-etal-2024-instruction}, Finance-LLaMa-8B \cite{tarun7r/Finance-Llama-8B}, Finance-R1 \cite{khazarai_personal_finance_r1}, LLaMa-RAG \cite{liu2024raginstructboostingllmsdiverse}. The third category features Vision-Language Models (VLMs), designed to process both images and text. For this group, we evaluated GPT-5, GPT-5-mini, GPT-5-nano, GPT-4.1, Mistral-3.1-24B \cite{jiang2023diego}, Gemma-3-12B \cite{gemma_2025}, LLaMa-3.2-11B \cite{grattafiori2024llama}, Sa2VA-8B \cite{sa2va}, Qwen-2.5-VL-7B \cite{qwen2.5-VL}, Qwen-2-VL-7B \cite{Qwen2VL}, Cosmos-7B \cite{azzolini2025cosmos}, and LLaVa-1.6-7B \cite{liu2023improved}. Finally, we assessed audio-language models, which handle audio and text inputs, including Voxtral-24B \cite{liu2025voxtral}, DeSTA-2.5-Audio-8B \cite{lu2025desta25Audio}, Qwen-2.5-Omni-7B \cite{xu2025qwen2}, Qwen-2-Audio-7B \cite{Qwen2-Audio}, Gemma-3n-E4B \cite{gemma_2025}.

\subsection{Benchmark Input Design}

To systematically evaluate the performance of different models, we designed a customized input and prompting strategy for each modality. 
%The objective was to create a standardized framework that ensures a fair comparison of baselines.
%while accommodating the unique input requirements of each model.

\paragraph{Text-Only Modality.} To ensure a fair evaluation across all baseline models, we standardized the input length to accommodate the smallest context window of 32K tokens. We set the threshold to 31K, because we need to reserve some space for the instruction prompt. As approximately 20\% of the transcripts in our dataset exceed this limit, we employed a targeted summarization strategy for these longer texts. Transcripts already under the 31K token threshold were left unmodified. To minimize information loss, we adopted a conservative, iterative summarization process as shown in Figure \ref{fig:summarization}. First, each transcript is segmented into chunks based on speaker type (e.g., Operator, Executive, Analyst). We then incrementally summarize the chunks that are least critical to the financial context, beginning with those from the ``Operator''. These sections typically contain procedural dialogue, such as call introductions and closings, rather than useful financial discussion. Every time a chunk is summarized, we recalculate the total token count, and the iterative process stops as soon as the transcript length falls below the 31K token threshold. If summarizing all ``Operator'' chunks is insufficient, the iterative process continues with chunks from ``Executives'' and then ``Analysts''. In practice, summarization usually stops midway through this process, leaving the majority of the transcript untouched and thus preserving as much original information as possible. Finally, the processed text was formatted into a single concatenated sequence for model input: ``\{Instruction\} \{Text Transcript\} \{Answer Format\}''. This structure provides the model with the complete verbal context of the conference call.

\noindent
For this summarization task, we utilized the BART-Large-CNN model \cite{DBLP:journals/corr/abs-1910-13461}, because it is lightweight, reliable, and does not require a specific instruction prompt to function effectively. For token number checking, we used ``tiktoken'' python library. While summarization was a necessary step for our experiments, we will release the original, full-length transcripts to support future research that may leverage models with larger context windows or more advanced summarization techniques.

\paragraph{Image-Text Modality.} Evaluating multi-modal models that accept text and images required a preprocessing step for the presentation slides, which were originally in PDF format. As most models do not support PDF inputs, we converted each page of a presentation into a separate image. To manage the image input limits common to these models, we developed a random sampling strategy. For each presentation, we randomly selected three images from the slide range, excluding the first and last pages, which typically contain non-substantive content like title or closing. In cases where a presentation had three or fewer content slides, all available slides were used. The final input for this modality was structured as: ``\{Images\} \{Instruction\} \{Text Transcript\} \{Answer Format\}''. Note that although our benchmark experiment uses only three images, we still release the full PDF in our dataset.

\paragraph{Audio-Text Modality.} For models capable of processing audio, we intentionally omitted the text transcript from the input. Since the audio recording contains the same verbal information as the transcript, this approach isolates the model's ability to comprehend and reason based on auditory signals alone, such as speaker tone and inflection. The input was therefore constructed as: ``\{Audio\} \{Instruction\} \{Answer Format\}''.

Across all modalities, the prompt structure was designed to test both the model's reasoning capabilities and its ability to follow instructions. We required the final answer to be placed in a specific, designated location within the output, which enabled reliable parsing for our automated evaluation pipeline. Detailed examples of the prompts for each modality are provided in the Appendix \ref{app:prompts}.

\subsection{Benchmark Metrics}

Since our benchmark involves only two possible outputs, accuracy naturally serves as a primary evaluation metric.
% :
% \begin{equation*}
% \text{Accuracy} = \frac{TP + TN}{TP + TN + FP + FN},
% \end{equation*}
% where $TP$ is True Positive, $TN$ is True Negative, $FP$ is False Positive, and $FN$ is False Negative.
However, accuracy alone can be misleading, particularly under class imbalance. To provide a more balanced evaluation, we report macro-averaged precision, macro-averaged recall, and macro-averaged F1-score. These metrics treat both positive and negative classes equally, regardless of their frequency in the dataset. 
% Below, we show the equation for Macro Precision:
% \begin{equation*}
% \text{Macro Precision} = \frac{\text{Precision}_{\text{pos}} + \text{Precision}_{\text{neg}}}{2},
% \end{equation*}
% where $\text{Precision}_{\text{pos}} = \frac{TP}{TP + FP}$ and $\text{Precision}_{\text{neg}} = \frac{TN}{TN + FN}$. We now show the equation for Macro Recall:
% \begin{equation*}
% \text{Macro Recall} = \frac{\text{Recall}_{\text{pos}} + \text{Recall}_{\text{neg}}}{2},
% \end{equation*}
% where $\text{Recall}_{\text{pos}} = \frac{TP}{TP + FN}$ and $\text{Recall}_{\text{neg}} = \frac{TN}{TN + FP}$. Finally, we show the equation for Macro F1-Score:
% \begin{equation*}
% \text{Macro F1-Score} = \frac{\text{F1}_{\text{pos}} + \text{F1}_{\text{neg}}}{2},
% \end{equation*}
% where $\text{F1}_{\text{pos}} = 2 \times \frac{\text{Precision}_{\text{pos}} \times \text{Recall}_{\text{pos}}}{\text{Precision}_{\text{pos}} + \text{Recall}_{\text{pos}}}$ and $\text{F1}_{\text{neg}} = 2 \times \frac{\text{Precision}_{\text{neg}} \times \text{Recall}_{\text{neg}}}{\text{Precision}_{\text{neg}} + \text{Recall}_{\text{neg}}}$.
\textit{By incorporating macro-averaged metrics alongside accuracy, we account for the dataset’s class imbalance and obtain a more reliable assessment of model performance across both positive and negative cases.}

\subsection{Implementation Details}

All experiments were conducted on a single NVIDIA A100 SXM4 GPU with 80GB of memory. For all baseline models, we maintained their official repository configurations and loaded them with bfloat16 precision to optimize computational efficiency. Beyond this precision setting, no other modifications were made to the models’ default parameters, ensuring a fair and reproducible comparison across all baseline evaluations.

\begin{table*}[ht]
\centering
\scalebox{0.75}{
\begin{tabular}{lcccccccccccccccc}
\toprule
\toprule
\multicolumn{1}{c}{} & \multicolumn{4}{c}{2019} & \multicolumn{4}{c}{2020} & \multicolumn{4}{c}{2021} & \multicolumn{4}{c}{Overall} \\
\multicolumn{1}{c}{} & Acc & Prec & Rec & F1 & Acc & Prec & Rec & F1 & Acc & Prec & Rec & F1 & Acc & Prec & Rec & F1 \\ \hline
\textit{General-Text} & \multicolumn{1}{l}{} & \multicolumn{1}{l}{} & \multicolumn{1}{l}{} & \multicolumn{1}{l}{} & \multicolumn{1}{l}{} & \multicolumn{1}{l}{} & \multicolumn{1}{l}{} & \multicolumn{1}{l}{} & \multicolumn{1}{l}{} & \multicolumn{1}{l}{} & \multicolumn{1}{l}{} & \multicolumn{1}{l}{} & \multicolumn{1}{l}{} & \multicolumn{1}{l}{} & \multicolumn{1}{l}{} & \multicolumn{1}{l}{} \\ \hline
GPT-oss (20B) & 0.73 & \cellcolor{bestbg}\textbf{0.57} & \cellcolor{bestbg}\textbf{0.59} & \multicolumn{1}{c|}{\cellcolor{bestbg}\textbf{0.58}} & 0.63 & 0.49 & 0.49 & \multicolumn{1}{c|}{0.48} & 0.75 & \cellcolor{bestbg}\textbf{0.52} & \cellcolor{bestbg}\textbf{0.52} & \multicolumn{1}{c|}{\cellcolor{bestbg}\textbf{0.52}} & 0.71 & \cellcolor{bestbg}\textbf{0.53} & \cellcolor{bestbg}\textbf{0.54} & \cellcolor{bestbg}\textbf{0.53} \\
Qwen-2.5 (14B) & \cellcolor{bestbg}\textbf{0.78} & 0.51 & 0.50 & \multicolumn{1}{c|}{0.49} & \cellcolor{bestbg}\textbf{0.77} & \cellcolor{bestbg}\textbf{0.52} & 0.52 & \multicolumn{1}{c|}{\cellcolor{bestbg}\textbf{0.52}} & \cellcolor{bestbg}\textbf{0.84} & 0.48 & 0.49 & \multicolumn{1}{c|}{0.48} & \cellcolor{bestbg}\textbf{0.80} & 0.51 & 0.50 & 0.50 \\
Mistral (7B) & 0.73 & 0.50 & 0.50 & \multicolumn{1}{c|}{0.50} & 0.68 & \cellcolor{bestbg}\textbf{0.52} & \cellcolor{bestbg}\textbf{0.53} & \multicolumn{1}{c|}{0.51} & 0.80 & 0.51 & 0.51 & \multicolumn{1}{c|}{0.51} & 0.74 & 0.51 & 0.51 & 0.51 \\
LLaMa-3.2 (3B) & 0.74 & 0.50 & 0.50 & \multicolumn{1}{c|}{0.50} & 0.75 & 0.51 & 0.51 & \multicolumn{1}{c|}{0.51} & 0.83 & \cellcolor{bestbg}\textbf{0.52} & \cellcolor{bestbg}\textbf{0.52} & \multicolumn{1}{c|}{\cellcolor{bestbg}\textbf{0.52}} & 0.78 & 0.51 & 0.51 & 0.51 \\
Gemma-3 (1B) & 0.70 & 0.43 & 0.45 & \multicolumn{1}{c|}{0.44} & 0.73 & 0.45 & 0.45 & \multicolumn{1}{c|}{0.45} & 0.77 & 0.45 & 0.45 & \multicolumn{1}{c|}{0.45} & 0.74 & 0.44 & 0.45 & 0.45 \\ \hline
\textit{Finance-FT} & \multicolumn{1}{l}{} & \multicolumn{1}{l}{} & \multicolumn{1}{l}{} & \multicolumn{1}{l}{} & \multicolumn{1}{l}{} & \multicolumn{1}{l}{} & \multicolumn{1}{l}{} & \multicolumn{1}{l}{} & \multicolumn{1}{l}{} & \multicolumn{1}{l}{} & \multicolumn{1}{l}{} & \multicolumn{1}{l}{} & \multicolumn{1}{l}{} & \multicolumn{1}{l}{} & \multicolumn{1}{l}{} & \multicolumn{1}{l}{} \\ \hline
Fin-LLaMa3 (8B) & 0.10 & 0.17 & 0.08 & \multicolumn{1}{c|}{0.09} & 0.12 & 0.23 & 0.07 & \multicolumn{1}{c|}{0.10} & 0.14 & 0.27 & 0.08 & \multicolumn{1}{c|}{0.12} & 0.12 & 0.22 & 0.08 & 0.11 \\
Fin-LLaMa (8B) & \cellcolor{bestbg}\textbf{0.70} & \cellcolor{bestbg}\textbf{0.46} & \cellcolor{bestbg}\textbf{0.47} & \multicolumn{1}{c|}{\cellcolor{bestbg}\textbf{0.46}} & \cellcolor{bestbg}\textbf{0.77} & \cellcolor{bestbg}\textbf{0.47} & \cellcolor{bestbg}\textbf{0.47} & \multicolumn{1}{c|}{\cellcolor{bestbg}\textbf{0.47}} & \cellcolor{bestbg}\textbf{0.77} & \cellcolor{bestbg}\textbf{0.46} & \cellcolor{bestbg}\textbf{0.46} & \multicolumn{1}{c|}{\cellcolor{bestbg}\textbf{0.46}} & \cellcolor{bestbg}\textbf{0.74} & \cellcolor{bestbg}\textbf{0.47} & \cellcolor{bestbg}\textbf{0.47} & \cellcolor{bestbg}\textbf{0.47} \\
Finance-R1 (1.7B) & 0.49 & 0.37 & 0.32 & \multicolumn{1}{c|}{0.34} & 0.48 & 0.42 & 0.32 & \multicolumn{1}{c|}{0.35} & 0.57 & 0.43 & 0.35 & \multicolumn{1}{c|}{0.38} & 0.52 & 0.40 & 0.33 & 0.36 \\
LLaMa-RAG (8B) & 0.08 & 0.14 & 0.05 & \multicolumn{1}{c|}{0.07} & 0.10 & 0.21 & 0.06 & \multicolumn{1}{c|}{0.09} & 0.12 & 0.26 & 0.08 & \multicolumn{1}{c|}{0.11} & 0.10 & 0.20 & 0.06 & 0.09 \\ \hline
\textit{Image-Text} &  &  &  &  &  &  &  &  &  &  &  &  &  &  &  &  \\ \hline
GPT-5 & 0.57 & \cellcolor{bestbg}\textbf{0.54} & \cellcolor{bestbg}\textbf{0.56} & \multicolumn{1}{l|}{0.51} & 0.52 & \cellcolor{bestbg}\textbf{0.54} & \cellcolor{bestbg}\textbf{0.59} & \multicolumn{1}{l|}{0.46} & 0.71 & 0.54 & \cellcolor{bestbg}\textbf{0.58} & \multicolumn{1}{l|}{0.53} & 0.62 & \cellcolor{bestbg}\textbf{0.55} & \cellcolor{bestbg}\textbf{0.58} & 0.52 \\
GPT-5-mini & 0.60 & 0.53 & 0.54 & \multicolumn{1}{l|}{0.52} & 0.53 & 0.52 & 0.54 & \multicolumn{1}{l|}{0.45} & 0.73 & 0.53 & 0.55 & \multicolumn{1}{l|}{0.52} & 0.64 & 0.53 & 0.55 & 0.51 \\
GPT-5-nano & 0.61 & 0.53 & 0.54 & \multicolumn{1}{l|}{0.52} & 0.48 & 0.52 & 0.54 & \multicolumn{1}{l|}{0.43} & 0.73 & 0.52 & 0.53 & \multicolumn{1}{l|}{0.52} & 0.62 & 0.52 & 0.54 & 0.50 \\
GPT-4.1 & 0.66 & 0.53 & 0.54 & \multicolumn{1}{l|}{\cellcolor{bestbg}\textbf{0.53}} & 0.57 & 0.53 & 0.56 & \multicolumn{1}{l|}{0.48} & 0.79 & 0.54 & 0.55 & \multicolumn{1}{l|}{\cellcolor{bestbg}\textbf{0.54}} & 0.69 & 0.53 & 0.55 & \cellcolor{bestbg}\textbf{0.53} \\
Mistral-3.1 (24B) & 0.72 & 0.49 & 0.50 & \multicolumn{1}{l|}{0.49} & 0.72 & 0.51 & 0.52 & \multicolumn{1}{l|}{\cellcolor{bestbg}\textbf{0.51}} & 0.86 & 0.52 & 0.51 & \multicolumn{1}{l|}{0.50} & 0.77 & 0.51 & 0.51 & 0.51 \\
Gemma-3 (12B) & 0.66 & 0.50 & 0.50 & \multicolumn{1}{l|}{0.50} & 0.65 & 0.52 & 0.54 & \multicolumn{1}{l|}{0.50} & 0.83 & \cellcolor{bestbg}\textbf{0.55} & 0.54 & \multicolumn{1}{l|}{\cellcolor{bestbg}\textbf{0.54}} & 0.73 & 0.52 & 0.53 & 0.52 \\
LLaMa-3.2 (11B) & 0.68 & 0.52 & 0.54 & \multicolumn{1}{l|}{0.50} & 0.49 & 0.52 & 0.54 & \multicolumn{1}{l|}{0.43} & 0.52 & 0.51 & 0.51 & \multicolumn{1}{l|}{0.47} & 0.58 & 0.52 & 0.54 & 0.48 \\
Sa2VA (8B) & 0.72 & 0.43 & 0.46 & \multicolumn{1}{l|}{0.44} & 0.77 & 0.45 & 0.46 & \multicolumn{1}{l|}{0.45} & 0.80 & 0.44 & 0.45 & \multicolumn{1}{l|}{0.45} & 0.76 & 0.44 & 0.46 & 0.45 \\
Qwen-2.5 (7B) & \cellcolor{bestbg}\textbf{0.78} & 0.52 & 0.51 & \multicolumn{1}{l|}{0.48} & \cellcolor{bestbg}\textbf{0.82} & 0.47 & 0.49 & \multicolumn{1}{l|}{0.47} & \cellcolor{bestbg}\textbf{0.88} & 0.45 & 0.49 & \multicolumn{1}{l|}{0.47} & \cellcolor{bestbg}\textbf{0.82} & 0.49 & 0.50 & 0.48 \\
Qwen-2 (7B) & 0.70 & 0.43 & 0.45 & \multicolumn{1}{l|}{0.44} & 0.71 & 0.49 & 0.48 & \multicolumn{1}{l|}{0.48} & 0.81 & 0.49 & 0.49 & \multicolumn{1}{l|}{0.49} & 0.74 & 0.47 & 0.47 & 0.47 \\
Cosmos (7B) & 0.75 & 0.44 & 0.48 & \multicolumn{1}{l|}{0.45} & 0.79 & 0.49 & 0.49 & \multicolumn{1}{l|}{0.49} & 0.86 & 0.47 & 0.49 & \multicolumn{1}{l|}{0.48} & 0.81 & 0.47 & 0.49 & 0.47 \\
LLaVa-1.6 (7B) & 0.68 & 0.51 & 0.51 & \multicolumn{1}{l|}{0.51} & 0.66 & 0.49 & 0.48 & \multicolumn{1}{l|}{0.47} & 0.84 & 0.54 & 0.53 & \multicolumn{1}{l|}{0.53} & 0.74 & 0.52 & 0.52 & 0.52 \\ \hline
\textit{Audio-Text} &  &  &  &  &  &  &  &  &  &  &  &  &  &  &  &  \\ \hline
Voxtral (24B) & 0.71 & 0.51 & 0.51 & \multicolumn{1}{l|}{\cellcolor{bestbg}\textbf{0.51}} & 0.69 & 0.51 & \cellcolor{bestbg}\textbf{0.52} & \multicolumn{1}{l|}{\cellcolor{bestbg}\textbf{0.50}} & 0.80 & \cellcolor{bestbg}\textbf{0.53} & \cellcolor{bestbg}\textbf{0.53} & \multicolumn{1}{l|}{\cellcolor{bestbg}\textbf{0.53}} & 0.74 & \cellcolor{bestbg}\textbf{0.52} & \cellcolor{bestbg}\textbf{0.52} & \cellcolor{bestbg}\textbf{0.52} \\
DeSTA-2.5 (8B) & 0.54 & 0.50 & 0.49 & \multicolumn{1}{l|}{0.47} & 0.52 & 0.50 & 0.49 & \multicolumn{1}{l|}{0.43} & 0.55 & 0.50 & 0.49 & \multicolumn{1}{l|}{0.43} & 0.54 & 0.50 & 0.49 & 0.45 \\
Qwen-2.5 (7B) & \cellcolor{bestbg}\textbf{0.78} & 0.46 & 0.50 & \multicolumn{1}{l|}{0.45} & \cellcolor{bestbg}\textbf{0.84} & \cellcolor{bestbg}\textbf{0.53} & 0.51 & \multicolumn{1}{l|}{\cellcolor{bestbg}\textbf{0.50}} & \cellcolor{bestbg}\textbf{0.88} & 0.51 & 0.50 & \multicolumn{1}{l|}{0.48} & \cellcolor{bestbg}\textbf{0.84} & 0.51 & 0.50 & 0.47 \\
Qwen-2 (7B) & 0.62 & 0.43 & 0.42 & \multicolumn{1}{l|}{0.43} & 0.69 & 0.46 & 0.44 & \multicolumn{1}{l|}{0.45} & 0.67 & 0.45 & 0.42 & \multicolumn{1}{l|}{0.43} & 0.66 & 0.44 & 0.43 & 0.43 \\
Gemma-3n (8B) & 0.35 & \cellcolor{bestbg}\textbf{0.53} & \cellcolor{bestbg}\textbf{0.53} & \multicolumn{1}{l|}{0.35} & 0.32 & 0.50 & 0.50 & \multicolumn{1}{l|}{0.31} & 0.33 & 0.48 & 0.46 & \multicolumn{1}{l|}{0.31} & 0.34 & 0.50 & 0.51 & 0.33 \\
\bottomrule
\bottomrule
\end{tabular}}
\caption{Model comparison on the FinCall-Surprise benchmark. The table is divided into four sections: General-Text refers to unimodal LLMs trained for general purposes; Finance-FT refers to unimodal LLMs finetuned on financial datasets; Image-Text refers to vision-language models that accept both images and text; and Audio-Text refers to audio-language models that accept both audio and text. For each metric and year, the highest score within each category is highlighted in \colorbox{bestbg}{\textbf{bold}}. (Prec, Rec, and F1 represent Macro Average Precision, Macro Average Recall, and Macro Average F1-score, respectively.)}
\label{tab:benchmark_result}
\end{table*}

\section{Benchmark Results}

We have listed all 26 models' result in Table \ref{tab:benchmark_result} including 5 general purpose text modality models, 4 financial finetuned models, 12 image-text modalities models, 5 audio-text modalities models. We have the following observations.

\subsection{High Performance Illusion}
\label{sec:illusion}

As shown in Table \ref{tab:benchmark_result}, many models appear to achieve strong results when evaluated only by accuracy. In fact, 15 out of 26 models surpass 70\% accuracy. However, this impression of high performance is misleading and illusional. Our dataset is heavily imbalanced toward positive samples, and when we examine precision, recall, and F1 score, most of the models fall below 55\%. This indicates that most models are biased toward predicting the majority (positive) class and fail to generalize to negative cases. 
% We believe that a well-trained model, given balanced and high-quality financial data, should be able to achieve both strong accuracy and unbiased performance, even when evaluated on imbalanced test sets. 
A plausible interpretation for this biased behavior is that the models have simply learned the statistical distribution of the financial domain, where positive outcomes are more frequent. This aligns with our own findings during the dataset construction phase, which confirmed that real-world financial data often exhibits significant imbalance.

A closer look also reveals that model accuracy tends to increase on the 2021 subset compared to earlier years. For instance, Qwen-2.5 in the General-Text category achieves accuracies of 0.78 and 0.77 in 2019 and 2020, respectively, but rises to 0.84 in 2021. However, this gain does not reflect genuine model improvement. Rather, it stems from the fact that the 2021 data is more imbalanced, with 89\% positive and only 11\% negative samples. When models default to predicting the majority class, their accuracy naturally rises under such conditions. This improvement is therefore illusory, as confirmed by its precision, recall, and F1 scores. For example, although Qwen-2.5 achieves 0.84 accuracy in 2021, its precision, recall, and F1 scores show no corresponding increase.

\subsection{Balanced Predictions and Robustness of Closed-Source Models}

Surprisingly, the four closed-source models evaluated, GPT-5, GPT-5-mini, GPT-5-nano, and GPT-4.1, achieved lower overall accuracy scores in the Image-Text category compared to smaller, open-source models. For instance, the highest accuracy among these closed-source models was 0.69, considerably lower than the 0.82 achieved by Qwen-2.5.
However, we argue that this accuracy metric is misleading (discussed in section \ref{sec:illusion}). A deeper analysis reveals that the GPT family, particularly GPT-5, consistently achieved the highest precision and recall scores nearly every year. This discrepancy arises from the models' prediction behavior. Manual inspection of the outputs showed that unlike many open-source models which predominantly default to a ``Positive'' label, the closed-source models generated a much more balanced distribution of ``Positive'' and ``Negative'' predictions. This avoidance of majority-class bias is what lowers their raw accuracy score. We interpret this as a positive sign, as it suggests the GPT models possess a more robust financial reasoning capability, allowing them to avoid the majority-class bias inherent in the dataset.
%Nonetheless, the inherent complexity of the earnings surprise prediction task remains a significant challenge for all models.

\subsection{Failure Analysis of Finetuned Financial Models}

While most models demonstrated reasonable performance on accuracy, a subset of models in the Finance-FT category performed exceptionally poorly. Notably, Fin-LLaMa3 and LLaMa-RAG achieved overall accuracy scores of only 12\% and 10\%, respectively. A qualitative analysis of their outputs revealed three primary failure reasons: response bias, poor instruction following, and degraded language generation.

First, unlike most other models, these two models show a strong tendency to predict ``negative'' labels. The reason for this bias is unclear, but it may stem from the nature of their finetuning data. Second, the models exhibit difficulty following instructions. To evaluate model performance systematically, our benchmark requires answers in a specific format: \texttt{Result = [[POSITIVE or NEGATIVE]]}. However, the finetuned models often failed to comply, producing outputs in inconsistent formats. Although we attempted to accommodate these variations, some responses could not be parsed reliably. We treated such cases as incorrect, since the ability to follow task instructions should be considered an essential component of performance. Third, in several instances the models failed to generate grammatically correct or meaningful sentences. We suspect that both the instruction-following failures and the degradation in basic language generation stem from over-finetuning on highly specialized financial datasets. In pursuit of domain-specific performance, finetuning may inadvertently over-penalize general language capabilities, diminishing the model’s broader usability. These findings highlight that when developing task-specialized models, it is important to ensure that this does not come at the expense of fundamental language competence and instruction-following ability.

\subsection{Do Image and Audio Inputs Improve Model Performance?}

As shown in Table~\ref{tab:benchmark_result}, model performance varies significantly across categories. Interestingly, we observe both modest improvements and notable degradations in accuracy when additional modalities are introduced. For instance, within the Qwen-2.5 family, the General-Text model with 14B parameters achieves an overall accuracy of 0.80. When image data is incorporated, the 7B Image-Text model improves to 0.82. With audio data, the 7B Audio-Text model increases the accuracy to 0.84. While these gains may appear modest, the fact that smaller multimodal models outperform their larger text-only counterpart suggests that visual and auditory inputs can provide complementary signals beyond the raw conference call transcripts.

However, in other cases, the addition of new modalities leads to performance degradation. For example, the text-only LLaMa-3.2 model with 3B parameters achieves a respectable accuracy of 0.78, yet its larger Image-Text variant with 11B parameters performs significantly worse, dropping to 0.58. Audio-Text models also exhibit highly inconsistent performance. While Voxtral achieves 0.74 accuracy and Qwen-2.5 achieves 0.84 accuracy, other models like DeSTA-2.5 and Gemma-3n reach only 0.54 and 0.34, respectively. These negative results suggest that, for many state-of-the-art models, supplementary signals from presentation slides and audio may introduce noise that cannot be effectively understood with textual content, ultimately hindering predictive capabilities. More broadly, current multimodal models often lack the robustness and consistency that text-only LLMs demonstrate when applied to single-modal financial data.

Taken together, our results highlight both the potential and the limitations of multimodal learning in this domain. On the one hand, certain models clearly benefit from the additional information that pure text alone cannot provide. On the other hand, many existing multimodal models cannot interpret complex financial signals across modalities. We anticipate that future research leveraging our multimodal benchmark will unlock deeper insights and enable models to capture information that plain transcripts alone cannot convey.

\section{Conclusions}

In this work, we introduced FinCall-Surprise, the first large-scale, open-source, multimodal benchmark for earnings surprise prediction, comprising synchronized word-level transcripts, full audio recordings, and presentation slides from more than 2,600 corporate conference calls between 2019 and 2021. Our evaluation of 26 state-of-the-art models highlights both the potential and the limitations of current approaches. We find that many open-source models achieve deceptively high accuracy due to class imbalance, while closed-source models generally provide more balanced predictions. Financial models fine-tuned on narrow datasets often lose basic instruction-following and language generation abilities, and existing multimodal models continue to struggle with effectively integrating image and audio in the financial domain. We release FinCall-Surprise to reduce reliance on proprietary data and provide an open, challenging benchmark for advancing models in earnings surprise prediction.

% By releasing FinCall-Surprise, we aim to reduce reliance on proprietary datasets and provide the community with an open, challenging benchmark that can drive the development of more robust multimodal architectures for earnings surprise prediction.

\clearpage
\section*{Limitations}
% \textcolor{red}{Add limitati}
A potential limitation of FinCall-Surprise lies in the natural class imbalance between positive and negative earnings surprises. Because positive surprises occur more frequently in real-world financial markets, our dataset inevitably reflects this biased distribution. While this imbalance makes the benchmark realistic, it also poses challenges for model evaluation. Future research that employs FinCall-Surprise for training could consider techniques such as upsampling, downsampling, or other modern data-balancing strategies to mitigate this limitation and better assess model robustness across both positive and negative cases.

\bibliography{custom}

\clearpage
\appendix

\section{Prompt Used in Our Experiments}
\label{app:prompts}

\subsection{Prompt Used in Text-Only Modality}

\begin{formal}
You are an expert equity research analyst.  Your task is to read the following earnings‐conference call transcript and decide whether next quarter’s earnings per share (EPS) will **beat** or **miss** consensus estimates. \newline

**Instructions**  \newline
1. **Label only**: respond with exactly one of:\newline
- `POSITIVE`  (indicating an expected positive EPS surprise)\newline
- `NEGATIVE`  (indicating an expected negative EPS surprise)  \newline
2. **Brief rationale** (1–2 sentences) explaining the key driver  \newline
3. **No other text**\newline

**Definitions**  \newline
- **POSITIVE**: indications of stronger‐than‐expected earnings or revenue growth, upbeat guidance, new partnerships, successful product launches, cost savings, market expansion, etc.  \newline
- **NEGATIVE**: indications of missed earnings or revenue declines, weak or withdrawn guidance, regulatory/legal setbacks, unexpected costs, competitive pressures, etc.\newline
    
**Transcript**:\newline
\textcolor{red}{\{transcript\}}\newline

**Answer format**  \newline
Rationale = [[ your rationale ]]  \newline
Result = [[ POSITIVE or NEGATIVE ]]\newline
\end{formal}

\subsection{Prompt Used in Image-Text Modality}

\begin{formal}
\textcolor{red}{[image\_1, image\_2, image\_3]}\newline
You are an expert equity research analyst. Your task is to read the following earnings‐conference call transcript and the accompanying slide deck images to decide whether next quarter’s earnings per share (EPS) will **beat** or **miss** consensus estimates.\newline

**Instructions**  \newline
1. **Label only**: respond with exactly one of:\newline
- `POSITIVE`  (indicating an expected positive EPS surprise)\newline
- `NEGATIVE`  (indicating an expected negative EPS surprise)  \newline
2. **Brief rationale** (1–2 sentences) explaining the key driver  \newline
3. **No other text**\newline

**Definitions**  \newline
- **POSITIVE**: indications of stronger‐than‐expected earnings or revenue growth, upbeat guidance, new partnerships, successful product launches, cost savings, market expansion, etc.  \newline
- **NEGATIVE**: indications of missed earnings or revenue declines, weak or withdrawn guidance, regulatory/legal setbacks, unexpected costs, competitive pressures, etc.\newline
    
**Transcript**:\newline
\textcolor{red}{\{transcript\}}\newline

**Answer format**  \newline
Rationale = [[ your rationale ]]  \newline
Result = [[ POSITIVE or NEGATIVE ]]\newline
\end{formal}

\subsection{Prompt Used in Audio-Text Modality}

\begin{formal}
\textcolor{red}{\{audio\}}\newline
You are an expert equity research analyst. Your task is to analyze the provided audio from an earnings conference call and predict whether next quarter’s earnings per share (EPS) will **beat** or **miss** consensus estimates.\newline

**Instructions**  \newline
1. **Label only**: respond with exactly one of:\newline
- `POSITIVE`  (indicating an expected positive EPS surprise)\newline
- `NEGATIVE`  (indicating an expected negative EPS surprise)  \newline
2. **Brief rationale** (1–2 sentences) explaining the key driver  \newline
3. **No other text**\newline

**Definitions**  \newline
- **POSITIVE**: indications of stronger‐than‐expected earnings or revenue growth, upbeat guidance, new partnerships, successful product launches, cost savings, market expansion, etc.  \newline
- **NEGATIVE**: indications of missed earnings or revenue declines, weak or withdrawn guidance, regulatory/legal setbacks, unexpected costs, competitive pressures, etc.\newline

**Answer format**  \newline
Rationale = [[ your rationale ]]  \newline
Result = [[ POSITIVE or NEGATIVE ]]\newline
\end{formal}

\section{Examples of Model Responses}
\label{app:example}

\noindent
\paragraph{GPT OSS 20B:}

\begin{formal}
<|channel|>analysis<|message|>We need to decide if next quarter EPS will beat or miss consensus. Transcript is about acquisition of Celgene by BMS and EPS accretion announced, many positive statements, expected near-term launches, synergy. But we need to see if EPS guidance they gave earlier? They said \"EPS guidance we issued this morning.\" But the details not present. However they mention strong EPS accretion day one, and near-term product launches. So likely positive. There's no mention of negative factors or earnings decline. Therefore predict positive EPS beat. Provide rationale: acquisition and near-term launches with high revenue potential and EPS accretion. Thus final answer: POSITIVE. Provide rationale in 1-2 sentences: e.g., acquisition creates immediate EPS accretion and multiple near 2011 term launches generating >\$15B potential, likely beating consensus.<|end|><|start|> assistant <|channel|> final <|message|> Rationale = [[The acquisition of Celgene is positioned to deliver immediate EPS accretion and the company expects six near 2011 term launches with over \$15B of peak sales, creating strong upside for earnings beyond current guidance.]]  Result = [[POSITIVE]]<|return|>
\end{formal}

\noindent
\paragraph{Qwen 2.5 14B:}

\begin{formal}
Rationale = [[ Alexion's earnings conference call indicates strong momentum with the launch of MG, successful pipeline advancements, including early approval for Ultomiris and promising data for NMO, and strategic acquisitions like Wilson Therapeutics and Syntimmune. The company plans to convert PNH patients to Ultomiris, expand into additional indications like ALS and PPMS, and maintain a focus on disciplined business development. These factors suggest robust growth prospects. Additionally, Alexion's emphasis on long-term sustainability and volume growth through strategic pricing aligns positively with investor expectations. ]] Result = [[ POSITIVE ]]
\end{formal}

\noindent
\paragraph{Mistral 7B:}

\begin{formal}
Rationale = Stronger-than-expected earnings growth due to the creation of a leading biopharma company with a broad and deep pipeline, leading marketed products portfolio, and multiple near-term product launches. Result = POSITIVE
\end{formal}

\noindent
\paragraph{LLaMa 3.2 3B:}

\begin{formal}
Result = POSITIVE  Rationale = The executives provided strong guidance on the upcoming launches of the combined company, including 6 near-term product launches with significant revenue potential. They also highlighted the pipeline's broad and deep nature, with a focus on long-term growth opportunities and the potential for value creation in the medium and long term.
\end{formal}

\noindent
\paragraph{Gemma 3 1B:}

\begin{formal}
Rationale = [[ The transaction represents a significant strategic realignment for Bristol Myers Squibb, positioning the company as a leader in oncology and immunology, leveraging the strengths of Celgene\u2019s pipeline and expanding its access to novel therapies. The combination offers substantial growth potential, with the combined company projected to deliver strong returns for shareholders.]] Result = [[ POSITIVE ]]
\end{formal}

\noindent
\paragraph{Finance-LLaMa3 8B:}

\begin{formal}
We'res how we see the company we see the product we see the product we see the company we see the company we see the company we see the company we see the company we see the company we see the company we see the company we see the company we see the company we see the company we see the company we see the company we see the company we see the company we see the company we see the company we see the company we see the company we see the company we see the company we see the company we see the company we see the company we see the company we see the company we see the company we see the company we see the company we see the company we see the company we see the company we see the company we see the company we see the company we see the company we see the company we see the company we see the company we see the company we see the company we see the company we see the company we see the company we see the company we see the company we see the company we see the company we see the company we see the company we see the company we see the company we see the company we see the company we see the company we see the company we see the company we see the company we see the company we see the company we see the company we see the company we see the company we see the company we see the company we see the company we see the company we see the company we see the company we see the company we see the company we see the company we see the company we see the company we see the company we see the company we see the company we see the company we see the company we see the company we see the company we see the company we see the company we see the company we see the company we see the company we see the company we see the company we see the company we see the company we see the company we see the company we see the company we see the company we see the company we see the company
\end{formal}

\noindent
\paragraph{Finance-LLaMa 8B:}

\begin{formal}
Assistant: POSITIVE. Explanation: next quarter 2019s earnings per share (EPS) will beat consensus estimates. The combination of Bristol-Myers Squibb and Celgene will create a leading biopharma company with strong potential for growth. The combined portfolio will have leading franchises in oncology, immunology and inflammation, and cardiovascular disease. The combined company will have a robust late-stage pipeline and early-stage pipeline across disease areas, as well as cutting-edge technologies and discovery platforms. The transaction is financially compelling on day one, with strong returns and immediate EPS accretion. The combined company will have significant growth opportunities, including 6 near-term product launches with more than \$15 billion in revenue potential. The combined company will have an industry-leading late-stage pipeline, with 6 near-term product launches, and a robust early-stage pipeline across disease areas. The combined company will also have cutting-edge technologies and discovery platforms that will enable the acceleration of new medicines for patients. The combined company will have strong potential for growth, with leading franchises in oncology, immunology and inflammation, and cardiovascular disease. The combined company will have significant growth opportunities, including 6 near-term product launches and a robust early-stage pipeline. The combination of Bristol-Myers
\end{formal}

\noindent
\paragraph{Finance-R1:}

\begin{formal}
<think> Okay, let's tackle this. The user wants to know if next quarter's EPS will beat or miss consensus estimates based on the earnings conference call transcript provided. First, I need to scan through the transcript for any mentions of EPS, revenue growth, cost savings, market expansion, or other factors that would indicate a positive or negative surprise. Looking at the transcript, the main points seem to be about the merger between Bristol-Myers Squibb and Celgene. They mention creating a leading biopharma company, combining their assets, and the strategic benefits. There are talks about expanding their portfolio, new product launches (like Revlimid, Pomalyst, and others), and the potential for growth in oncology and immunology areas. The executives talk about the financial implications of the deal, stating that they'll deliver strong returns with immediate EPS accretion. They also mention the combined pipeline with potential revenue from 6 near-term product launches. There's a lot of emphasis on the synergy and the long-term growth opportunities. However, the transcript doesn't mention any actual EPS numbers or revenue figures for the current quarter. The focus is more on the strategic moves and future outlook rather than past or current EPS performance. The key drivers here are the strategic alignment, combined pipeline potential, and the expected revenue growth from new products. The definition of POSITIVE is when earnings or revenue growth is stronger than expected, or there's new partnerships, successful product launches, cost savings, etc. The transcript mentions new product launches, expanded portfolio, and the potential for growth in key areas. Since these are positive factors, it's likely that the EPS will beat expectations. The NEGATIVE would be if earnings or revenue declines are expected, but the transcript doesn't mention any declines. The focus is on growth and positive outcomes. So the key driver here is the strategic expansion and the potential for strong revenue from the new product launches and expanded portfolio, which would lead to a positive EPS surprise. </think> Rationale = [The transcript emphasizes strategic alignment, expanded portfolio, and high-growth potential from combined pipelines, including 6 near-term product launches. These factors suggest strong revenue and EPS growth, indicating a positive surprise.]  Result = [POSITIVE]
\end{formal}

\noindent
\paragraph{LLaMa-RAG:}

\begin{formal}
Good a good an the the the and **Fees to have a good a positive to and  to ** to the  (and to come the positive (best (1 the negative  The. We the ** to have an overall to come and **to the positive the 2018 to come the **the same to come the ** to come the best of the 2018 and ** to come the best of the ** (Positive the positive the next year we the to a good a good (the negative and the negative the 1 ** the positive the **to the **the. We **the negative the higher the positive the positive (the (negative. We the 1.  **the positive (Best of a positive the  we, **the positive (1 **the positive and **the same the **the negative to come the positive the positive (1 **the negative the negative the positive (negative Following the The  The  We the negative the best (the positive (The positive, positive ( negative the positive, the negative The positive  to come the Positive ** to come the positive the positive, and Positive the positive, positive (negative the negative the following an overall the
\end{formal}

\end{document}